\documentclass[preprint,proceedings]{rmaa}
\usepackage{paralist}

\usepackage{psfrag,color}

\newcommand{\ltsima}{\mbox{$\; \buildrel < \over \sim \;$}}
\def \simlt{\lower.5ex\hbox{\ltsima}}            % < over ~
\def \gtsima{\mbox{$\; \buildrel > \over \sim \;$}}
\def \simgt{\lower.5ex\hbox{\gtsima}}            % > over ~

\newcommand{\K}{\hbox{{\rm K}}}
\newcommand{\s}{\hbox{{\rm s}}}

\newcommand{\kms}{\hbox{{\rm km s$^{-1}$}}}

\newcommand{\lya}{{\mbox Ly$\alpha$}}
\newcommand{\Cfour}{{\rm C\,{\sc IV}}}
\newcommand{\Nfive}{{\rm N\,{\sc V}}}

\newcommand{\Osix}{{\rm O\,{\sc VI}}}
\renewcommand{\H}{\ion{H}{1}}

\newcommand{\Hep}{\ion{He}{2}}

\newcommand{\h}{{\rm H\,{\sc I}}}

\newcommand{\hep}{{\rm He\,{\sc II}}}
\newcommand{\hepp}{{\rm He\,{\sc III}}}

\newcommand{\mtau}{{\hbox{$\bar\tau_{\rm eff}$}}}

\SetYear{2002}
\SetConfTitle{Galaxy Evolution: Theory and Observations}

\title{Numerical simulations of the intergalactic medium}

\author{
  Tom Theuns\altaffilmark{1,2} 
	}	

\altaffiltext{1}{Institute of Astronomy, University of Cambridge, UK}
\altaffiltext{2}{Universitaire Instelling Antwerpen, Antwerpen, Belgium}

\shortauthor{Theuns}
\shorttitle{Simulations of the intergalactic medium}

\fulladdresses{
\item Tom Theuns: Institute of Astronomy, Madingley Road, Cambridge CB3
0HA, UK and Universitaire Instelling Antwerpen, Universiteitsplein
1, B-2610 Antwerpen, Belgium.
}
\listofauthors{T. Theuns}
\indexauthor{Theuns, T.}

\abstract{The intergalactic medium at redshifts 2--6 can be studied
observationally through the absorption features it produces in the
spectra of background quasars. Most of the UV-absorption lines arise in
mildly overdense regions, which can be simulated reliably with current
hydrodynamical simulations. Comparison of observed and simulated spectra
allows one to put contraints on the model's parameters.}

\resumen{El medio intergal\'actico a redshifts 2--6 se puede estudiar
observacionalmente gracias a los rasgos de absorci\'on que produce en el
espectro de los cuasares de fondo de campo. La mayori\'a de las lineas de
absorci\'on UV aparecen en regiones ligeramente sobrepobladas, que pueden
ser simuladas de manera fiable con las simulaciones hidrodin\'amicas
actuales. La comparaci\'on del espectro observado con el simulado nos
permite poner limites a los par\'ametros del modelo.}

\addkeyword{Cosmology: observations}
\addkeyword{Cosmology: theory}
\addkeyword{Galaxies: formation}
\addkeyword{Intergalactic medium}
\addkeyword{quasars: absorption lines}

\begin{document}
% Typeset article header
\maketitle

\section{General}
\label{sec:intro}

Spectra of quasars contain hundreds of absorption lines, which arise in
the mostly smoothly distributed intergalactic medium (IGM) along the
line of sight. Most of the lines are due to the absorption by neutral
hydrogen, which produces the quasar's \lq Lyman-$\alpha$ forest\rq{}
(Lynds 1971; see Rauch 1998 for a recent review), but transitions of
other elements such as \Cfour{}, \Nfive{} and \Osix{} are seen as well
(e.g. Cowie et al.{} 1995). 

A fully neutral IGM would block all quasar light below the \lya-
transition, and so the fact that some flux is observed implies that the
IGM is very highly ionized (Gunn \& Peterson 1965; Bahcall \& Salpeter
1965). At redshifts $z\ltsima 3$, quasars produce enough ionizing
photons to explain the inferred high ionization levels (e.g. Rauch et
al.\ 1997), but their dwindling numbers at higher redshifts suggest
that galaxies must make-up an increasingly important contribution to
the background at higher $z$. The nature of the sources responsible for
reionizing the universe is still being debated, as is the epoch of
reionization. The recent observations of redshift $z\sim 6$ quasars
with a significant stretch of \lya-forest with zero flux to within the
noise (Becker et al.\ 2001; Djorgovski et al.\ 2001), are consistent
with a detection of the transition from a neutral to a highly ionized
IGM.

Hydrodynamical simulations of hierarchical structure formation in a
cold dark matter (CDM) dominated universe, have been very successful in
reproducing the statistical properties of the observed \lya-absorption,
in the redshift range $0\ltsima z\ltsima 4$ (see e.g. Efstathiou,
Schaye \& Theuns 2000 for a recent review). These simulations show that
the weaker \lya-lines are predominantly produced in the filamentary and
sheet-like structures that form naturally in this cosmology. These
structures are at modest densities $0.3\ltsima \rho/\langle\rho\rangle
\ltsima 10$, and so can be reliably simulated, and semi-analytical (Bi
\& Davidsen 1997) and analytical models (Schaye 2001) provide valuable
insight into the dominant processes that shape the lines.

The ionizing radiation photo-heats the gas, and establishes a
density-temperature relation
$T=T_0(\rho/\langle\rho\rangle)^{\gamma-1}$ (Hui \& Gnedin 1997). The
gas temperature $T_0(z)$ and the exponent $\gamma(z)$ retain a memory
of the reionization history, because the thermal time-scales are long
in the low-density IGM (Miralda-Escude\'e \& Rees 1994; Haehnelt \&
Steinmetz 1998). The widths of the \lya-lines can be used to measure
$T_0$ and $\gamma$ (Schaye et al.{} 1999). Wavelets provide another way
to characterize changes in line-widths (Theuns \& Zaroubi
2000). Finally, the mean level of absorption depends on $T_0$, and this
can be exploited as well to search for a sudden change in $T_0$ which
could arise from the epoch of \Hep{} reionization (Miralda-Escud\'e \&
Rees 1994; Theuns et al.{} 2002c).

The metals detected in the IGM were presumably synthesized in stars,
and later expelled into the surroundings by galactic winds. This
process is currently not well understood, and difficult to simulate
(see e.g. Mac Low \& Ferrara 1999), because of the complexity of the
interstellar medium in galaxies (McKee \& Ostriker 1977). Preliminary
simulations of such winds indicate that they can indeed pollute the IGM
with enough metals to reproduce the data, without significantly
changing the properties of the hydrogen absorption lines (Theuns et
al.{} 2002d).

Section~2 briefly summarizes the current status of IGM simulations,
section~3 discusses constraints on reionization and section~4
illustrates the effects of galactic winds.

\section{Simulations}
\label{sec:simulations}
Structure formation in a CDM-dominated cosmology is characterized by a
set of numbers, namely the density of matter, vacuum-energy and
baryons, the Hubble constant $h$, the amplitude of the power spectrum,
and the helium abundance. Recently, most simulations have used small
variations around a popular set of $(\Omega_m,\Omega_bh^2, h,
\sigma_8,Y)$=(0.3,0.019,0.65,0.9,0.24), and assumed a geometrically
flat universe.

Once these numbers are chosen, the linear power spectrum can be computed
(e.g. using {\sc cmbfast}, Seljak \& Zaldarriaga 1996), and a
representation of a random realization of such a density field using
particles can be generated using the Zel'dovich (1970) approximation.
The equations of motion are then integrated into the non-linear regime
(see e.g. Efstathiou et al.{} 1985). 

Baryons can be added in a variety of ways, using Smoothed Particle
Hydrodynamics (e.g. Hernquist et al.{} 1996; Haehnelt \& Steinmetz
1996; Wadsley \& Bond 1996; Theuns et al.{} 1998), or finite-difference
schemes (e.g. Cen \& Ostriker 1992; Zhang, Anninos \& Norman 1995;
Bryan et al.{} 1999). In addition, the evolution of the
ionizing-background needs to be specified, either estimated from the
simulation itself, or computed separately (e.g. Haardt \& Madau 1996)
and imposed by hand. Given a UV-background, the photo-ionization
equations determine the abundances of ionized hydrogen, helium (and
other species). Non-equilibrium effects are important during
reionization (Abel \& Haehnelt 1999), and need to be modeled
separately.

Mock spectra are computed along random sight lines through the
simulation box (e.g. Theuns et al.{} 1998), which can be made to look
like observed low or high-resolution spectra by convolving with a
Gaussian to mimic a given spectral resolution, and adding noise (see
Theuns, Schaye \& Haehnelt 2000 for details). These mock spectra are
then analyzed with the same tools as the data, e.g. using {\sc
vpfit}\footnote{http://www.ast.cam.ac.uk/\~{}rfc/vpfit.html} (Webb
1987) to fit Voigt profiles to the absorption features.

\section{The thermal history of the IGM and reionization}
\begin{figure}
  {\includegraphics[bb=5 200 500 600,clip,width=0.99\columnwidth]{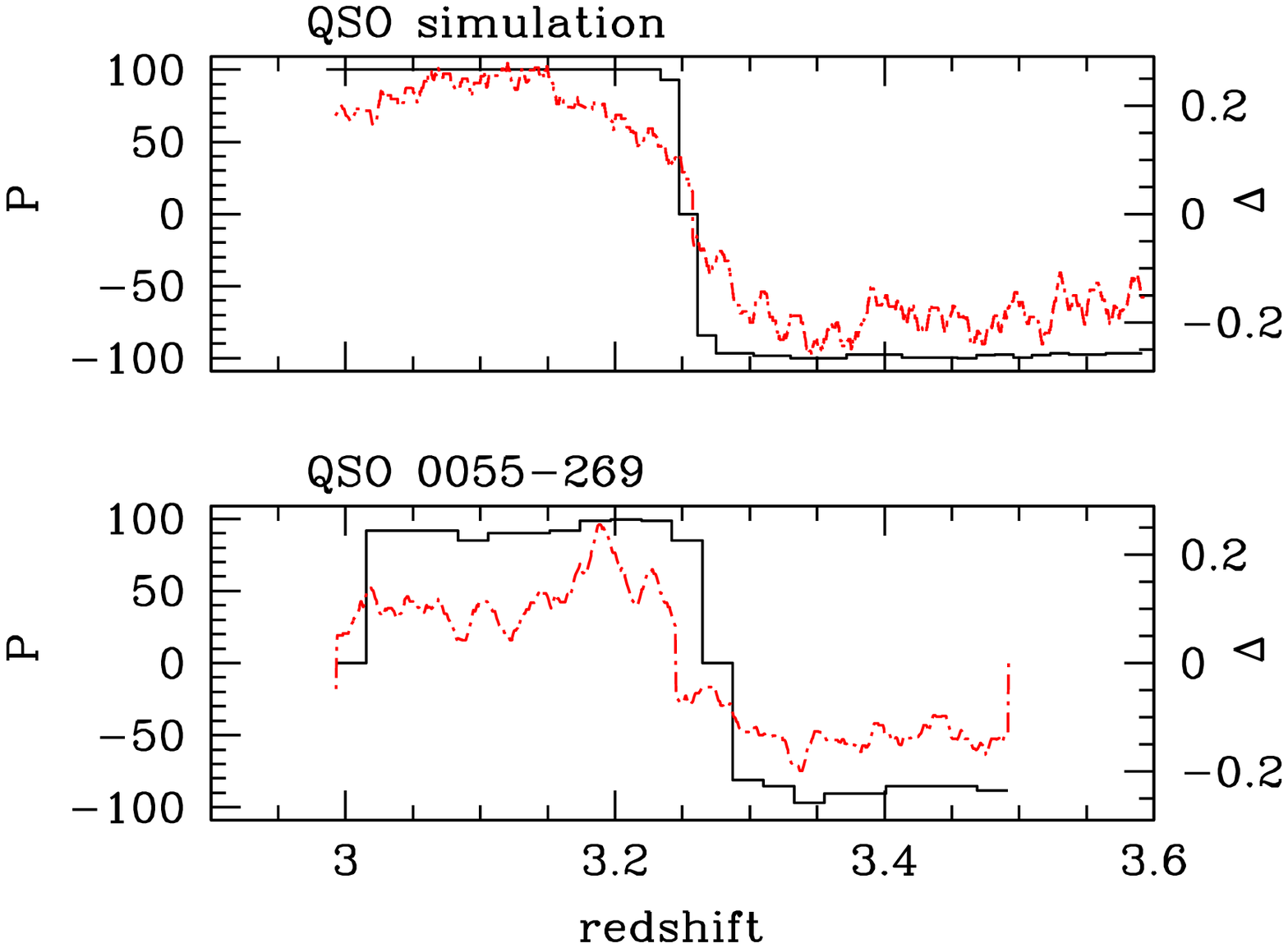}}
  \caption{Application of the wavelet method to a simulated spectrum (top
panel) and QSO~0055--269 (bottom panel). In the simulations, the low
redshift half is 50 per cent hotter than the high redshift half. This
jump is detected at the 99.5 per cent level. A similar jump is seen in
the observed spectrum.}
  \label{fig:wvlt}
\end{figure}

\begin{figure}
  {\includegraphics[bb=10 100 600
  500,clip,width=0.95\columnwidth]{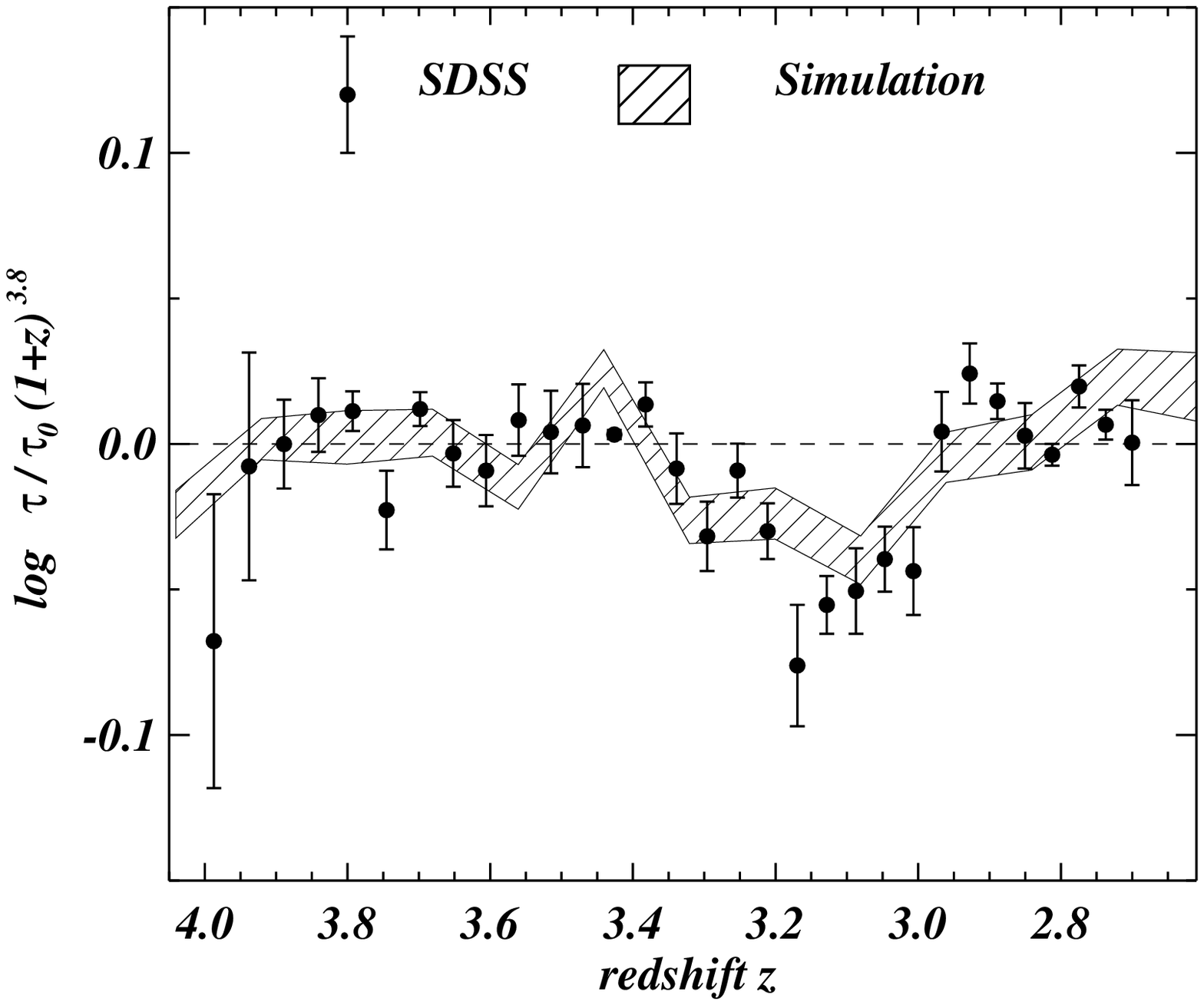}} \caption{Deviation of
  the effective optical depth from a power-law evolution,
  $\mtau/(1+z)^{3.8}$, for the SDSS data smoothed on 3000\kms (symbols
  with error bars) and a hydrodynamical simulation of a $\Lambda$CDM
  model, in which \Hep\ reionization starts at $z=3.4$ (hashed
  region). The temperature increase associated with \Hep\ reionization
  causes \mtau\ to drop below the power-law evolution in the
  simulation. This characteristic dip matches the feature
  detected in the SDSS data very well .}  \label{fig:sloan}
\end{figure}

In the simulations, the line-widths of the hydrogen \lya-lines depend
on temperature, in the sense that thermal broadening introduces a
cut-off in the line-widths for given column-density (Schaye et al.{}
1999). Such a cut-off is also clearly seen in observational samples
(Kirkman \& Tytler 1997). The relation between temperature and cut-off
can be calibrated with simulations, and hence the thermal history can
be reconstructed (Schaye et al.{} 2000; Ricotti, Gnedin \& Shull 2000;
Bryan \& Machacek 2000; McDonald et al.{} 2001). Schaye et al.{} (2000)
found evidence for an increase in $T_0$ around redshift $z\sim 3$,
which they interpreted as evidence for \Hep{} reionization. Normal
stars do not produce photons that are hard enough to ionize \Hep{}, and
so the reionization epoch for \H{} and \Hep{} can be quite different,
if \H{} reionization is due to galaxies.

Theuns et al.{} (2002a) used wavelets as basis functions to characterize
\lya-forest spectra. The amplitude of a narrow wavelet is a measure of
the typical widths of the absorption features, and so a change in
temperature -- as might result from \Hep{} reionization -- would lead
to a change in rms amplitude of the wavelet. Wavelets have the
advantage over Voigt profile fitting that the wavelet spectrum of a
given signal is unique, since wavelets form an orthogonal basis. In
addition, the decomposition into wavelets is extremely
fast. Figure~\ref{fig:wvlt} shows a measure of the wavelet amplitude
for a simulated spectrum in which $T_0$ increases by 50 per cent at
redshift $z=3.3$ (wavy line), and for the spectrum of quasar~0055--269.
There is strong evidence for a sharp increase in $T_0$ in the observed
spectrum. The significance of a given amplitude is denoted by the
histogram-like full line (see Theuns et al.{} 2002a for details). The
significance level of the temperature change is more than 99 per cent.

Such a temperature change will also influence the mean absorption
$\exp(-\mtau)$ in the \lya-forest. This is because the neutral hydrogen
fraction is determined by the balance between photo-ionizations and
recombinations, and the recombination coefficient depends on
temperature. Figure~\ref{fig:sloan} compares the predicted evolution of
the effective optical depth, \mtau{}, with the evolution measured from
the Sloan Digital Sky Survey by Bernardi et al.{} (2002). The good
agreement suggests that \Hep\ re-ionization has been detected in the
SDSS data set (Theuns et al.{} 2002c).

\begin{figure}[!t]
  {\includegraphics[bb=0 150 600
  700,clip,width=0.95\columnwidth]{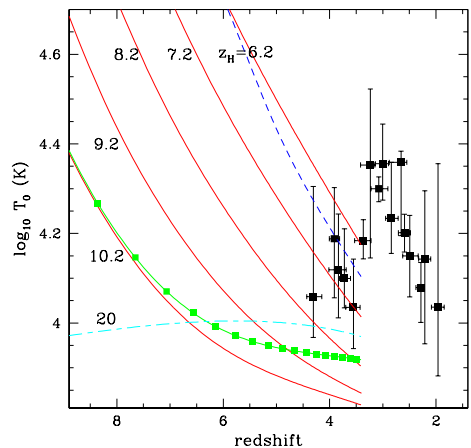}} \caption{The
  temperature evolution of the IGM above redshift 3.4. Symbols with
  error bars are determined from QSO observations by Schaye et al.{}
  (2000). The solid curves indicate the evolution of the temperature at
  the mean density for various assumed \h\ reionization redshifts
  $z_{\rm H}$, as indicated. The post-hydrogen reionization temperature
  is assumed to be $T_0=6\times 10^4~\K$ and the hydrogen
  photo-ionization rate is $\Gamma_\h=10^{-13}~ \s^{-1}$ (the short
  dashed line has $\Gamma_\h=10^{-14}~ \s^{-1}$). The \hep\
  photo-ionization rate is adjusted so that the \hepp\ abundance
  $x_\hepp\approx 0.1$ at $z=3.5$. The solid line connecting filled
  squares is for $z_{\rm H}=10.2$, and a higher \hep\ photo-ionization
  rate, $x_\hepp(z=3.5)=0.6$. Finally, the long dashed line has $z_{\rm
  H}=20$, but a still higher \hep\ photo-ionization rate,
  $x_\hepp(z=3.5)=0.95$.  If, as expected, He is mostly singly ionized
  at $z\ga 3.5$, then the rapid decrease in $T_0$ after reionization
  places an upper limit of $z_{\rm H}< 9$ on the redshift of hydrogen
  reionization.}  \label{fig:reion}
\end{figure}

\begin{figure*}[!t]
  \includegraphics[width=\textwidth]{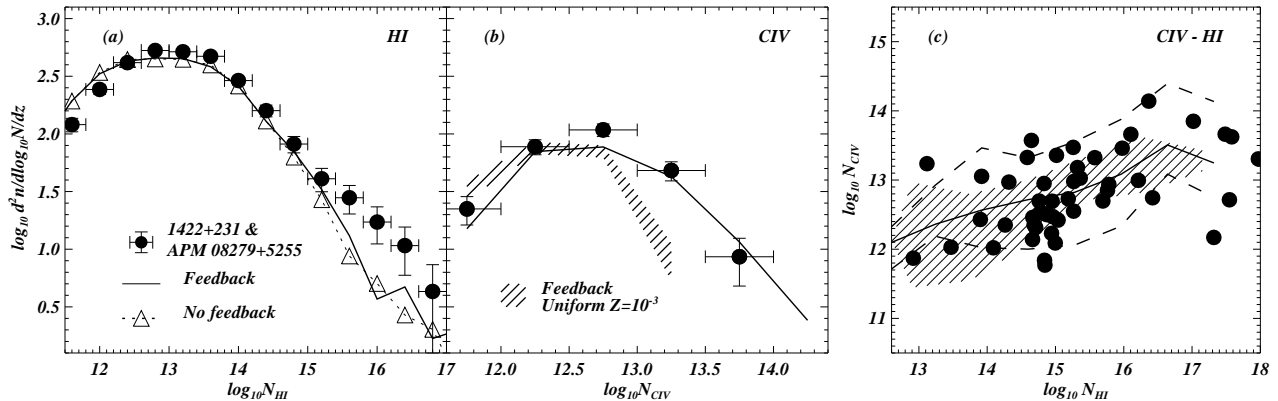}%
  \hspace*{\columnsep}%
	\caption{Column density distribution functions
  (CDDFs) of \H\ (panel a), \Cfour\ (panel b), and the \Cfour{} versus
  \H{} column density of systems (panel c). Filled circles refer to the
  combined line-lists of quasars Q1422+231 (Ellison et al.\ 2000;
  $z_{\rm em}=3.6$) and APM 08279+5255 ($z_{\rm em}=3.91$), and the
  solid line indicates the results for the feedback simulation at
  $z=3$. Panel (a): The \H\ CDDF of the feedback simulation is nearly
  identical to that of the simulation without feedback (dotted line and
  triangles), and both fit the observations very well for $\log N_{HI}
  \la 15$. Panel (b): The feedback simulation matches the observed
  \Cfour\ CDDF, but imposing a uniform metallicity ($Z=10^{-3}Z_\odot$,
  hashed region) does not work as well. Panel (c): \Cfour{} versus \h{}
  column density, for systems identified on a smoothing scale of $\pm
  150\kms${}. The median for the feedback simulation is indicated with
  the full line, with dashed lines indicating the 5 and 95 per
  centiles. The hashed region is the corresponding range for the
  uniform metallicity case. The feedback simulation reproduces the
  observed median and scatter better than the uniform $Z$ case.}
  \label{fig:metals}
\end{figure*}

If this interpretation is correct, then the high values of $T_0$
inferred above $z\sim 4$, imply that hydrogen reionized relatively late
(Theuns et al.{} 2002b). Figure~\ref{fig:reion} illustrates how $T_0$
drops rapidly after \h{} reionization, and how this can be used to
constrain the \h{} reionization epoch to be below a redshift of 9. If
\h{} reionization happened at $z>9$, then $T_0$ would have
decreased (due to adiabatic expansion) below the values measured around
$z\sim 4$.

\section{Metals in the IGM}
\label{sect:metals}

Feedback from star formation is thought to play an important role in
the formation of galaxies. Observations of star-bursts, both at low and
high redshift, show evidence for strong galactic winds with a mass-loss
rate comparable to the star-formation rate. Such winds could be
responsible for enriching the IGM with metals. Most strong \lya-lines
have associated \Cfour{} absorption, and at least at lower $z$, \Osix{}
as well (Carswell, Schaye \& Kim 2002). At present it is not clear
whether only regions close to galaxies are metal enriched, or if most
of the IGM contains metals.

It has been suggested that these metals may have been distributed by an
early generation of population~III stars at very high redshifts. One of
the arguments for this pre-enrichment scenario, is the suspicion that
galactic winds at lower redshifts, if they were sufficiently volume
filling to produce the required pollution with metals, would destroy
the filaments that produce the \lya-forest.

We have performed simulations of the IGM in order to investigate this
question (Theuns et al.{} 2002d). In these simulations, star-formation
generates hot regions that expand in the form of a metal enriched
wind. This feedback implementation strongly quenches the star-formation
rate. Crucially, there is almost no difference in the properties of the
\lya-lines between this simulation and a similar simulation without
feedback. This is because the hot bubbles prefer to expand into the
lower density surroundings of the galaxies, thereby leaving the
filaments that produce most of the \H{} lines intact, and because the
volume filling factor of the winds is small. The metal enriched gas
produces about the observed number of \Cfour{} lines, and the same
large scatter between \H{} and \Cfour{} column-density as is observed
(Fig.~\ref{fig:metals}). It will be interesting to examine in more
detail whether the statistics of the simulated metal lines agree with
the data.

\section*{Acknowledgments}
I am grateful to my collaborators to allow me to present our results,
and thank PPARC for the award of an Advanced Fellowship. I also thank
the organizers of the meeting for a very enjoyable week. This work was
supported by the European Community Research and Training Network \lq
The Physics of the Intergalactic Medium\rq{} and was conducted in
cooperation with Silicon Graphics/Cray Research utilizing the Origin
2000 super computer at the Department of Applied Mathematics and
Theoretical Physics in Cambridge.

\end{document}